# An EMG-based Eating Behaviour Monitoring System with Haptic Feedback to Promote Mindful Eating

Ben Nicholls, Chee Siang Ang, Kanjo Eiman, Panote Siriaraya, Woon-Hong Yeo, *Member, IEEE*, and Athanasios Tsanas, *Member, IEEE*

*Abstract* – **Mindless eating, or the lack of awareness of the food we are consuming, has been linked to health problems attributed to unhealthy eating behaviour, including obesity. Traditional approaches used to moderate eating behaviour often rely on inaccurate self-logging, manual observations or bulky equipment. Overall, there is a need for an intelligent and lightweight system which can automatically monitor eating behaviour and provide feedback. In this paper, we investigate: i) the development of an automated system for detecting eating behaviour using wearable Electromyography (EMG) sensors, and ii) the application of such a system in combination with real time wristband haptic feedback to facilitate mindful eating. Data collected from 16 participants were used to develop an algorithm for detecting chewing and swallowing. We extracted 18 features from EMG and presented those features to different classifiers. We demonstrated that eating behaviour can be automatically assessed accurately using the EMG-extracted features and a Support Vector Machine (SVM): F1-Score=0.94 for chewing classification, and F1-Score=0.86 for swallowing classification. Based on this algorithm, we developed a system to enable participants to self-moderate their chewing behaviour using haptic feedback. An experiment study was carried out with 20 additional participants showing that participants exhibited a lower rate of chewing when haptic feedback delivered in forms of wristband vibration was used compared to a baseline and non-haptic condition (F (2,38) = 58.243, p <0.001). These findings may have major implications for research in eating behaviour, providing key new insights into the impacts of automatic chewing detection and haptic feedback systems on moderating eating behaviour with the aim to improve health outcomes.**

*Index Terms*—Eating behaviour monitoring; Haptic feedback; Mindful eating; Mobile and wearable devices.

## I. Introduction

According to a report from the U.S. Department of Labour, the average person spends 1.18 hours a day eating [1]. Oftentimes, during eating people may engage in additional concurrent activities such as working, driving or reading. By engaging in concurrent activities, people become arguably less aware of the extent of time they devote to eating. This *mindless eating* – or the lack of awareness of the food we are consuming – has been linked to the obesity epidemic and other health problems attributed to unhealthy eating behaviour [2, 3]. For example, the speed of food consumption has been associated with increased Body Mass Index (BMI) [4], diabetes [5], and various eating disorders [6]. Hence, investigating eating behaviour interventions may have wide ranging implications including weight management and eating disorder treatment.

Self-reporting and reflection are often considered important activities to facilitate behaviour change [7]. Such activities can help maintain a state of 'mindful' eating, which is important to counter automatic eating and environmental influences [3], and facilitate reflection upon behaviour change goals. Current studies looking into eating speed often rely on participant self-monitoring or manual observation in experimental settings. Alternative approaches to studying eating speed have made use of a *mandometer*, an electronic scale measuring the weight of food over time, to estimate intake rate [6]. Although such tools provide an objective measure of eating speed, they do not provide sufficient and detailed evaluation of eating processes such as chewing and swallowing.

The focus of this work is therefore twofold: i) Study 1 focuses on the development of an automated system for detecting eating behaviour (chewing and swallowing) using Electromyography (EMG) signals; ii) Study 2 aims to investigate the feasibility of using haptic feedback using a smart wristband to facilitate mindful eating using the detection technique developed in Study 1.

The paper is structured as follows. In section II, we provide a review of relevant literature. Section III focuses on Study 1: the development of an algorithm for chewing and swallowing detection. Section IV presents an experiment examining the effectiveness of a haptic feedback system for mindful eating (Study 2). The discussion of the experimental results and

Manuscript submitted on 24 July 2019. This work was supported in part by the Engineering and Physical Sciences Research Council (UK).

N. P. Nicholls is with the School of Engineering and Digital Arts, University of Kent, UK (e-mail: bpn2@kent.ac.uk)
C.S. Ang is with the with the School of Engineering and Digital Arts, University of Kent, UK (e-mail: csa8@kent.ac.uk).
E. Kanjo is with the School of Science and Technology, Nottingham Trent University, UK (e-mail: eiman.kanjo@ntu.ac.uk)
P. Siriaraya is with the School of Information and Human Science, Kyoto Institute of Technology, Japan (e-mail: spanote@kit.ac.jp)
W Yeo is with George W. Woodruff School of Mechanical Engineering and the Wallace H. Coulter Department of Biomedical Engineering at Georgia Institute of Technology. (e-mail: whyeo@gatech.edu)
A. Tsanas is with the Usher Institute, University of Edinburgh, UK (e-mail: Athanasios.Tsanas@ed.ac.uk).

implications for future work is provided in section V and VI.

## II. RELATED WORKS

### A. Links between eating rate and health

Previous studies have investigated the effect of eating rate on food intake quantity through controlled experiments. For example, Kokkinos et al. [8] conducted a study using timed eating period and food quantity to control eating speed, and measured hunger stimulating and inhibiting hormone levels in the blood. They reported higher concentration of hunger reducing hormones after a slower meal and hypothesised that this might indicate eating rate could be related to overconsumption of calories. Similarly, Zhu and Hollis [9] investigated the effect of controlled chewing thoroughness (chew count) and found that increased chewing thoroughness was associated with reduced eating rate and food palatability. Zandian et al. [6] compared *linear eaters* (people who eat at a constant rate) and *decelerated eaters* (people who slow down during the meal) during eating sessions with intake speeds where feedback was provided. They found that participants in the decelerated eating group demonstrated difficulty maintaining set eating speeds. Ioakimidis et al. [10] conducted a similar study, evaluating the effect of feedback upon the eating rate of linear eaters and people with eating disorder. They reported that changing the eating rate of linear eaters resulted in similar consumption patterns as those identified in people with eating disorder. This suggests that susceptibility to external influences may put linear eaters at risk of eating disorders, and that eating rate feedback may be a useful intervention tool to assist people to achieve the desired eating pattern [10].

Various diverse health factors may be related to chewing rate, directly or indirectly. For instance, Yamazaki et al. [11] examined 6,827 participants and concluded that masticatory performance and eating rate can be considered a potential risk factors and are associated with diabetes. There have also been studies suggesting a link between eating rate and 'stress-eating'. Adam and Epel [12] reported that those who release a large amount of cortisol in response to stress consumed more calories following application of high stress tests. Tasaka et al. [13] built on these hypotheses, relating salivary cortisol levels to chewing rate after study sessions involving stress loading and chewing at different rates, reporting reduced cortisol levels after fast chewing. Collectively, these studies concluded that there may be an association between psychopathological stress responses and eating behaviour, and also that chewing faster might contribute to stress relief.

### B. Limitations of current techniques in logging eating behaviour

The two main approaches for tracking eating are self-logging, and through manual observation (i.e. observations by human raters). Self-reported measures offer an easy approach to log diet for tracking eating disorders or weight management [7], or for large scale population studies of eating behaviour [14]. However, such measures are intrinsically subjective and might be unreliable or prone to bias [15]. For instance, in a large study of 4,808 participants to compare self-reported and clinically measured height and weight, Spencer et al. [16] reported overestimated height and underestimated weight. Similar effects were shown in other studies [17], and such bias was also observed during reliability assessments of eating disorder screening questionnaires [18]. The main limitation of manual observation-based studies is time and resource demands, which restricts the amount of recorded data one can analyse. In any large-scale study, the collection of high-quality data is time consuming and requires considerable resources. For example, Bajic [19] conducted a study of the effects of music on eating amongst 103 participants, which involved manual analysis of approximately 52 hours of video footage. Other studies overcome similar issues through strict experimental protocols to simplifying recorded data [20]. Some automated solutions exist, such as using a *mandometer*, or automated systems of eating behaviours. However, these approaches are relatively restricted in purpose and are immobile, thus limiting their applicability in practical settings.

### C. Using mobile technology to promote healthy eating

Over the last few years many studies highlighted advantages of mobile technologies in promoting healthy eating, i.e. the ability of mobile devices to provide users with an easily accessible platform which enables convenient recording of data regarding eating behaviours, and receiving relevant feedback about their dietary choices (see [21]). Examples of such systems include mobile based calorie monitoring systems which have been used in both scientific research and commercial context to encourage users to modify their dietary intake and consume food according to their dietary requirements. Notable examples include an image-based mobile food recording system, which uses before-after photographs of foods and beverages consumed by users [22]. Such technology has also been sought to help in managing specific diseases where dietary monitoring plays a key role (such as in diabetes care where eating habits are monitored in combination with physical activity to help patients manage their blood glucose levels) [23].

Whilst prior applications tend to focus on improving lifestyle-based eating behaviours, researchers have also examined how improvements during the eating process might lead to potential health benefits. For example, the eating rate has been the focus of many studies, particularly in relation to factors such as obesity [14] and diabetes risk [11], and has been suggested as a potential target for behavioural change [14]. Recently, the concept of *mindful eating* has been proposed as a technique to help regulate eating rate behaviour [2]. Mindfulness is the psychological process of bringing one's attention to experiences occurring in the present moment. Since eating is generally considered as a type of automatic behaviour, we have a tendency to consume food without conscious consideration. By helping people maintain a state of mindfulness during eating and more consciously examining hunger and satiation, individuals may be able to better "override" automatic eating behaviours. Such techniques have been proposed as an intervention to help fight against obesity [2]. However, in order to effectively monitor and provide feedback about eating behaviours, most existing studies tend to rely on users manually entering details about their food consumption which as discussed previously, requires considerable effort and could be prone to bias and participant error. Hence, for a behavioural change system to be useful, practical and unobtrusive, a monitoring technique would need to be employed to allow real-time monitoring of eating rate.

Mobile technologies have been proposed as a low-cost way to measure eating rate [24]. Jasper et al. [25] implemented an automated system for monitoring bite rate based upon hand motion captured by a wrist worn gyroscope which was evaluated under controlled and real-life conditions. They reported that feedback reduced the number of bites, but that this resulted in compensatory behaviour permitting increased intake [17]. The use of automated EMG-based eating detection for the monitoring of eating rate is another viable alternative [26]. Prior EMG studies approach the detection of eating rate through the detection of chewing activities, by using signal thresholds to identify periods of signal activity which denote rhythmic chewing events. Chewing is typically represented in EMG signals of the masticatory muscles by a burst of signal amplitude, occurring in a rhythmic sequence throughout the course of eating. The onset and termination of muscle activity is generally determined through the use of a predefined threshold; identifying onset and termination as the points at which the signal crosses the given threshold value. However, this approach has been found to be an unreliable approach, prone to false positives [27]. In addition, EMG signal activity of many facial muscle groups may be sensitive to inter-muscular cross-talk, where the detected activity might not be associated with underlying chewing aspects that we want to be characterising.

## III. STUDY 1: AUTOMATED DETECTION OF EATING BEHAVIOURS USING EMG

In this study, we developed an algorithm aiming to provide accurate and robust detection of chewing and swallowing events using EMG signals.

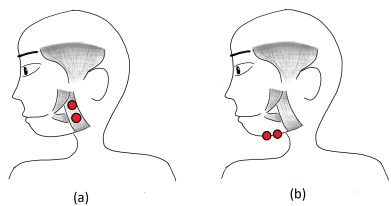

Figure 1: Surface electrode placement positions for EMG measurement of the (a) Masseter muscles, and (b) Suprahyoid muscles, based on [30].

### A. Data collection

The data collection system consisted of custom hardware and software paired with a physiological sensor device and a standard laptop computer. Participants were mounted with standard surface electrode sensors (#H124SG, Covidien, Ireland) connected to a Bluetooth enabled EMG measurement and transmitter unit (Shimmer 3, Shimmer Sensing, Ireland). All data was collected with a sampling frequency of 1024 Hz. Chewing and swallowing activities were monitored using EMG signals. For the purpose of mastication, the two primary masticatory muscles groups are the 'masseter muscles' and the 'temporalis muscles' used predominantly to control the elevation of the mandible [29]. In the context of EMG, Criswell and Cram [28] demonstrate the similarity of the signals from the two sites during chewing; describing mastication as the predominant action identifiable from the masseter muscles, and "assistance in chewing" as an important action of the temporalis. The masseter has also been described as easy to identify and reliable, which is a valid consideration for the purpose of reproducibility of this work [30]. The approximate position of electrode placement is presented in Figure 1.

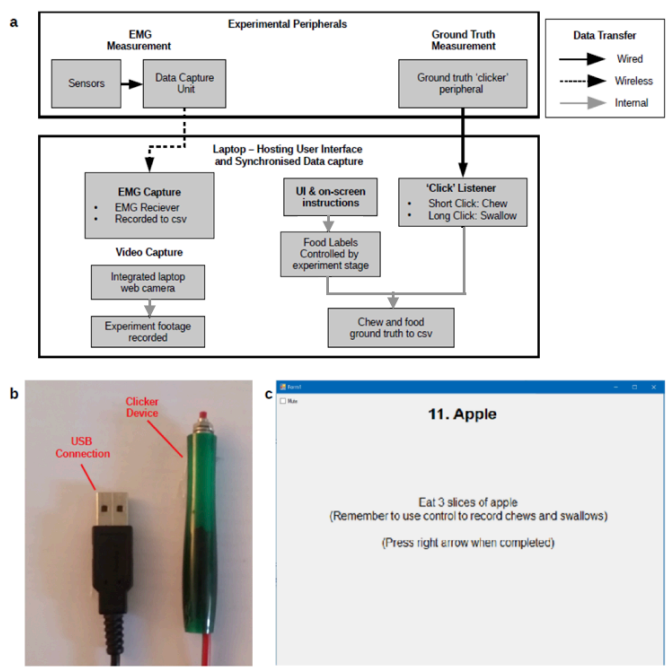

Figure 2: Components of the data capture system. a) a flowchart summarising experimental data collection system. b) shows the custom 'clicker' peripheral connected to a laptop via USB connection, to permit ground truth self-logging. c) shows example of experimental application screen, displaying instructions for data collection (in conjunction with audio instructions).

The data collection software was developed using the C# .Net platform and the Shimmer API[1]. Participants were able to self-report individual chews and swallows by performing a short click or long-hold of a 'clicker' device respectively (see figure 2b). All data was recorded concurrently and was synchronised. Video footages of the participants were captured to complement ground truth recording via the 'clicker'. In addition, the software also served to guide participants through the data collection, providing textual and verbal instructions (see figure 2c). Approval for the data collection procedures was granted by University of Kent Faculty of Sciences Research Ethics Advisory Group for Human Participants (Ref No 0721718).

---

[1] https://github.com/openmhealth/shimmer

## B. Participants

To generate the data, 16 participants were recruited from a research university (details of the University are not provided to protect participants' anonymity). Participants were selected to include a range of physical attributes (age, gender, height and weight). Overall, the age of the participants was between 18-40. Nine of the participants were female. Seven were considered to be overweight (BMI > 25) and one was considered slightly underweight (BMI=18.1). Participants were provided with a range of food items to consume. We selected five different food types which were representative of the textures and viscosities found in different food categories: apple, jam sandwich, pizza, yoghurt, and water. Participants were each asked to consume 18 portions of each food item, over the two iterations of the experimental procedure and the various meal sections. Each solid food item was cut into small bite-size portions, approximately 2.5cm square in the case of pizza and sandwiches, and apple slices 2 cm by 2.5 cm. Yoghurt was provided in a small container along with a 5 ml spoon. A portion of yoghurt was defined as a single spoonful. Unlimited water was provided and a portion was described to participants as a small mouthful.

Participants were asked to follow on-screen instructions guiding them through the experimental procedure: 5 minutes of baseline measurement, 5 minutes speaking aloud, head motion, and consumption of a small meal. Head motion was also carried out at times while eating to simulate normal movement during eating. Inclusion of reading and head motion was to permit training of classifiers which are robust to unrelated activity. Following completion, the sensors were removed from the participants, replaced, and the procedure was repeated. This process was followed to mitigate effects where minor changes in sensor placement might adversely impact the quality of data recorded. Each participant recorded two data sets, however for 3 of these participants only one dataset was considered viable due to hardware faults, and one participant elected not to return to take part in a session. Overall, a total of 28 datasets were collected, each comprising approximately 20 minutes of EMG data recorded during a combination of activities and food consumption. We processed 384 minutes of data from 16 participants. This includes 5 minutes of sitting still and 5 minutes of speech which were collected during each session. The remainder of the data consisted of participants consuming a small meal. During eating, a total 16,237 eating events were recorded, 14,180 chews and 2,057 swallows. The distribution of food labels is shown in Table 1.

## C. Data processing and feature extraction

The data was filtered and processed to eliminate noise and movement artefacts. Specifically, a unidirectional Butterworth bandpass filter was applied to the EMG signal with a low cut-off frequency at 20 Hz and a high cut-off frequency at 500 Hz (with cut-off order 5). The signal was then rectified using a full wave digital rectifier and normalised such that values lay within the 0-1 range. Each dataset was collected with self-reported ground truth labels (See Table 1). Whilst this gave a good indication of individual chew and swallow events, it was only an approximate indicator of the signal activity ground truth and did not guarantee the identification of uniform and predictable onset and termination times. To correct this, the ground truth for each dataset underwent automatic and manual review to ensure fidelity. Firstly, automatic correction of chewing event onset and termination was applied, using threshold based activity detection (based on the EMG of the masseter muscle). Accurate ground truth timings could then be identified, where periods of potential EMG activity intersect or lay within close temporal proximity to ground truth timestamps and used to correct ground truth. The same process was repeated for swallow ground truth correction, using submental activity. However, as these muscles also exhibited some activity during chewing, manual review of swallow EMG activity and video footage was used to confirm swallow ground truth onset and termination. The threshold value (*thr*) for this was determined using the approach suggest by Abbink et al. [31] and Li et al. [32] for EMG detection:

$$thr = \mu_0 + j * \delta_0 \qquad (1)$$

where $\mu_0$ is the mean of the baseline, $\delta_0$ is the standard deviation of the baseline, and $j = 5$.

TABLE 1
THE NUMBER OF EATING EVENTS RECORDED FOR EACH FOOD TYPE (N=16)

| Number of Recorded Eating Events | | | |
|---|---|---|---|
| **Type of Food** | **Class label** | | **Total** |
| | **Chew** | **Swallow** | |
| Apple | 3,595 | 369 | 3,964 |
| Sandwich | 4,282 | 376 | 4,658 |
| Pizza | 6,073 | 395 | 6,468 |
| Yoghurt | 230 | 330 | 560 |
| Water | 0 | 587 | 587 |
| **Total** | 14,180 | 2,057 | 16,237 |

Given that swallowing typically spans a longer period of time compared to chewing, we decided to treat this as binary classification problems: i) chewing classification - where all activities (including non-eating activities) were considered NA apart from chewing; ii) swallowing classification - where all activities were considered NA apart from swallowing. We down-sampled the EMG signal by a factor of 10 and computed features using a sliding, overlapping hamming window, which we set to 0.5 sec (512 samples) for chewing and 1.625 sec (1664 samples) for swallowing.

Features were extracted from the two signal channels and the sample was labelled according to a period of inactivity (NA), or a chew (C) or a swallow (S) event. A total of 18 features were extracted across two channels of EMG and used in the classification models based on previous literature (for details see the S1 in the Supplementary Material). In addition, there were considerable imbalances in the class labels in the final test and training sets, towards the inactive class. To compensate for this imbalance, class weights were computed and applied to the training data during training of all models. Class imbalances in the test sets were also liable to cause anomalous results during testing. To account for this the test sets were resampled at testing, down-sampling the majority classes to match the minority.

*D. Statistical mapping and model validation*

We used different classifiers to assess binary differentiation of chewing and swallowing events: Support Vector Machine (SVM) with linear kernel, Linear Discriminant Analysis (LDA), Decision Tree (DT), and Extra Trees meta estimator (ET). The statistical models' performance was assessed on a random selection of 25% of participants (4 participants). Furthermore, a leave-one-subject-out evaluation technique was employed, where in each run we trained the model using the samples from the k-1 participants and testing on the data from the k-th participant. The hyper-parameters of the classifiers were tuned using k-fold cross validation (k = 3) with grid search. The model performance was assessed using precision, recall, and the F1-score which are widely used in binary class classification settings. The latter provides a good compromise between the sensitivity and specificity of a statistical model.

TABLE 2
PERFORMANCE SUMMARY FOR THE CHEWING AND SWALLOWING CLASSIFIER BASED ON THE LEAVE ONE OUT EVALUATION METHOD

| Average F1-score related metrics for each class | | | |
|---|---|---|---|
| Class | Precision | Recall | F1-score |
| Chewing (C) | 0.95 | 0.95 | 0.95 |
| Swallowing (S) | 0.87 | 0.88 | 0.87 |
| Average | 0.91 | 0.92 | 0.91 |

TABLE 3
F1-SCORE FOR THE LEAVE ONE PARTICIPANT OUT

| The F1-score based on each test case | | | | | |
|---|---|---|---|---|---|
| Test case Number | F1-score per Classifier model | | Demographics | | |
| | Chew | Swallow | Age Range | Gender | BMI |
| 1 | 0.95 | 0.89 | 18-25 | Female | 25.00 |
| 2 | 0.96 | 0.88 | 26-35 | Female | 21.00 |
| 3 | 0.95 | 0.81 | 36-45 | Male | 24.30 |
| 4 | 0.96 | 0.86 | 26-35 | Male | 20.00 |
| 5 | 0.97 | 0.92 | 18-25 | Female | 25.50 |
| 6 | 0.92 | 0.94 | 26-35 | Female | 25.95 |
| 7 | 0.93 | 0.87 | 18-25 | Female | 25.97 |
| 8 | 0.94 | 0.86 | 18-25 | Female | 25.00 |
| 9 | 0.95 | 0.86 | 18-25 | Female | 22.28 |
| 10 | 0.94 | 0.83 | 18-25 | Female | 34.21 |
| 11 | 0.97 | 0.95 | 26-35 | Male | 27.00 |
| 12 | 0.97 | 0.80 | 18-25 | Male | 20.07 |
| 13 | 0.98 | 0.88 | 26-35 | Male | 18.08 |
| 14 | 0.94 | 0.82 | 26-35 | Male | 36.16 |
| 15 | 0.93 | 0.86 | 26-35 | Male | 20.32 |
| 16 | 0.91 | 0.91 | 18-25 | Female | 30.00 |
| Average | 0.95 | 0.87 | | | |
| SD | 0.02 | 0.04 | | | |

Table 2 shows the results for the leave-one-participant-out evaluation. F1-score for the evaluation of the model with each test case is shown in Table 3. Overall, there was a low standard deviation between test cases for the F1-score for both models, with a deviation of only 0.02 for the chewing classifier and 0.04 for the swallowing classifier. This low standard deviation of the F1-scores support the conclusion that the models generalise well to entirely unknown participants. Furthermore, the variation in age, gender, and BMI value across the participants suggest that these factors have little impact on the detection and classification of EMG signals during eating. For chewing, no test cases reported an F1-score of under 0.91 and the high scoring cases for chewing prediction (with F1-score above 0.96) were found to be evenly distributed between high BMI and normal BMI.

## IV. STUDY 2: REAL TIME HAPTIC FEEDBACK FOR MINDFUL EATING

In Study 2, an experimental study was carried out to investigate the effectiveness of our proposed haptic feedback system. To achieve this, we first adapted the eating detection algorithm from Study 1 to work in real time. Then, a mobile application was developed integrating this real time algorithm and haptic feedback using a smart wristband.

*A. Development of Real Time Chewing Detection Algorithm*

The previous section focused on the post-hoc classification of swallowing and chewing activity after data had been collected and pre-processed. The feedback system developed in Study 2 required real-time, or near real-time, detection of chewing events which could then be used to extrapolate information regarding chewing rate and providing feedback. The same dataset used in section III was used in the training and testing of the new live chewing detection algorithm. Since we are interested in the chewing activity, only the EMG channel corresponding to the masseter muscle was used. This helped minimise participant's exposure to unfamiliar sensors on their face which were potential distractors for the feedback study.

We computed features for each signal segment of 0.5 seconds: the mean of the signal for each segment, the standard deviation, maximum amplitude, root mean square value, integrated EMG, mean frequency, and mean frequency band power. The features were normalised using reference voluntary contractions to determine the appropriate maximal amplitude expected during eating. The reference amplitude was obtained during a short period of calibration (through eating one piece of each of the available foods) for each participant, during which the reference values were calculated. Afterwards, each entry in the final feature array was labelled as either occurring during a burst of EMG activity related to chewing behaviour (C) or as inactivity or unrelated activity (NA). A linear SVM based model was then trained using the available data. As for the hyper-parameters, a penalty value of 5 (through cross validation grid search) and a squared hinge loss function was used. For testing purposes, leave-one-participant-out approach was used. To compensate for class imbalances, the test sets were re-sampled at testing, down-sampling the majority classes to match the minority. The model was then evaluated based on the F1-Score, Precision and Recall.

Table 4 provides a summary of the evaluation results. Overall, for the classification of chewing activity in a real-time scenario from single channel EMG, the model resulted in an average Recall, Precision and F1-Score all of 0.92. Although these results demonstrate a small loss in performance from the model developed in the previous section, this loss was not substantial enough to suggest any detrimental impact resulting from the real-time approach to signal processing.

TABLE 4
THE AVERAGE PERFORMANCE OF THE REAL-TIME CHEW CLASSIFICATION MODEL BASED ON THE LINEAR SVM ALGORITHM. (USING LEAVE-ONE-PARTICIPANT-OUT)

| Class Label | Precision | Recall | F1-Score |
| --- | --- | --- | --- |
| N/A | 0.90 | 0.94 | 0.92 |
| Chew | 0.94 | 0.89 | 0.92 |
| Avg/Total | 0.92 | 0.92 | 0.92 |

### B. Implementation of the Real-Time Chewing Detection and Haptic Feedback System

The system consists of a Bluetooth enabled EMG signal capture device (Shimmer 3) connected to standard surface electrodes (#H124SG, Covidien, Ireland) axed across the masseter muscles on the dominant side of the user. To demonstrate the applications capacity in a mobile context, the measured signal was streamed live via Bluetooth to a smartphone (Samsung Galaxy S6) running Android version 3.0. The smartphone receives signal via a custom application and acts as a local intermediary between the signal capture device and remote classier, and also handled user feedback regarding chewing rate.

Figure 3 provides an overview of the chewing detection, monitoring, and feedback system. A laptop (Dell, Inspiron 75594) connected to the mobile device via Bluetooth connection acted as a remote server with custom software handling signal processing and classification. It calculated chewing rate information, permitted live monitoring and returned live chewing rate information to the phone for feedback. Feedback was delivered via a Microsoft Band device (Microsoft Band 2).

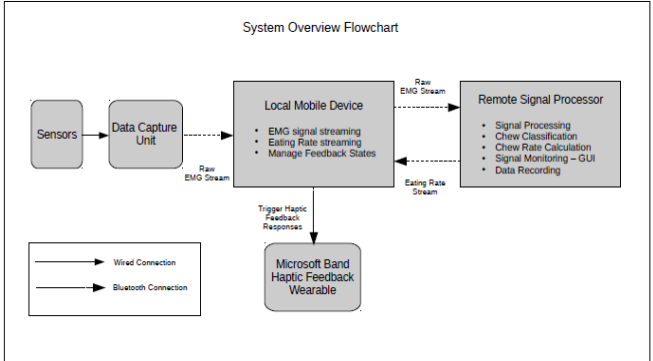

Figure 3: An overview of the chewing detection and Haptic Feedback system developed in this study

### C. Signal Processing and Classification Software

A custom application hosted on the laptop server was developed to process and classify the incoming data. The application was developed using Python 2.7 and TKinter, with matplotlib modules used for the purpose of providing a graphical user interface and visualising the live signal. The Linear SVM was implemented using the *Sci-kit* learn python library. Chewing bursts were estimated using a voting filter over a small window (of 8 samples), to reduce the probability of unexpected and individually occurring false positives.

The classification model was designed to return a positive prediction for all samples classified as occurring during an EMG chew burst. This enabled the system to determine the chewing events, which is defined as the period occurring between the onset and termination of positive predictions. Upon the termination of each detected chewing event, the predicted label, timestamps of the onset and termination of the event as well as the time duration of each event were logged on an output file. The chewing rate was then calculated based on the onset and termination timestamps of each detected chewing event (calculated as the number of chews per second) using Eq. (2):

$$CR = \frac{1}{n}\sum_{i=0}^{L} f(chew_{event}); f(x) = \begin{cases} 1, & \text{if } x_{onset} \geq (t-n) \text{ and } x_{term} \geq t \\ 0, & \text{otherwise} \end{cases}$$
(2)

Giving the average number of chews per second over the last $n$ seconds, where $chew\_event_i$ is a chewing event occuring during the session, $L$ is the number of chewing events observed during the session, $x_{onset}$ is the starting time of chewing event $x$, $x_{term}$ is the termination point of chewing event $x$, $t$ is the current time, and $n=5$. As the approximate duration of chewing events has been identified as 0.5 seconds in our earlier experiments, it was possible to measure chewing rate over the last 5 seconds with the timings of approximately 5 chews. This was deemed to provide acceptable accuracy for chewing rate calculation while attempting to minimise the time error for feedback response.

### D. Haptic Feedback System

Whilst biofeedback systems often make use of visual or audio feedback, visual feedback was disregarded for this study as it would require special attention while eating. Audible feedback on the other hand would be overtly obvious to other individuals in social scenarios. Therefore, in this study, we explored the use of haptic feedback. Haptic feedback is able to provide relatively covert feedback that would not demand special attention whilst still acting to draw the attention of the users back to their eating behaviour. The Microsoft band was configured to provide four different patterns of vibrations based on a normalised eating rate (between 0.0-1.0): (1) No haptic pulses (representing a 'low' eating rate of around 0.0-0.3), (2) Periodic individual haptic pulses (representing a 'moderate' eating rate of around 0.3-0.6), (3) Periodic double haptic pulses (representing a 'high' eating rate of around 0.6-0.8), and (4) High intensity double haptic pulses (representing a 'very high' eating rate of around 0.8-1.0).

### E. Experimental Study

A within-participant study was carried out to determine the effects of real-time feedback provided by our system on short term eating behaviour. Each participant was asked to participate in three different conditions where (1) in the control condition, they were asked to eat normally, (2) in the none-feedback condition, they were asked to self-moderate their eating rate, and (3) in the haptic feedback system, they were asked to self-moderate their eating rate using our proposed haptic system. The hypothesis of our experiment study is as follows: (1) the haptic feedback system would result in a reduced chewing rate in comparison to the none feedback and control conditions and

(2) the haptic feedback system would provide participants with more awareness in regards to the self-moderation of the eating rate in comparison to the none feedback and control conditions. Approval for the experimental procedures was granted by University of Kent Faculty of Sciences Research Ethics Advisory Group for Human Participants (Ref No 0721718).

*1)* Participants

20 additional participants were recruited from a research university (details of the University are not provided to protect participants' anonymity) (aged 18-50, 10 female). Only healthy participants were recruited with no dietary restrictions to the foods provided for the study. Consent was obtained to record anonymised sensor data and survey responses as well as to the record audios of the interview. The majority of participants were within a healthy weight range according to their BMI (13), whilst 3 were found to be slightly underweight (BMI less than 18.5), and 4 were found to be overweight (BMI greater than or equal to 25).

*2)* Materials

The system specified in the previous section was used during the course of this study. Participants had adhesive electrode sensors axed over their masseter muscles, following the placement procedure outlined in Figure 1, and were equipped with a Microsoft Band 2 for the study duration. The smartphone and remote processing laptop, which were included as part of this system, were placed nearby, but out of line of sight of the participants. The food selection was duplicated from previous data collection methodology involved in the development of chewing classification algorithms (section III).

*3)* Study Process

Each participant took part in a single study session consisting of three phases: a control phase involving unrestricted normal eating, and two treatment phases involving self-moderation of the eating rate, with and without feedback. At the beginning of the session, participants were equipped with the sensing equipment. Participants were then presented with food allotted to them for the experiment and asked if they would like to make any substitutions or reductions (participants ate the same type of food for all conditions which they participated in). Afterwards, the food was divided into three portions for each phrase of the study. Participants took part in the three phases of eating, completely consuming one portion of food during each phase. In the first phrase (the control condition), participants were asked to eat the food normally. This phrase served both to help assess the normal eating performance of each participant and allowed our software to be calibrated. Following this, participants were asked to self-moderate their eating rate based on two conditions:

- *Self-moderation eating without Haptic Feedback condition (No-Feedback):* In this condition, participants were asked to attempt to moderate their eating rate, trying to estimate their normal eating speed and slow down while eating the provided food portion.

- *Self-moderated eating, with haptic feedback (Feedback):* In this condition, participants were asked to moderate their eating behaviour with the help of our haptic feedback system. A brief training session was carried out at the beginning of this phase in which the chewing rate haptic feedback system was then demonstrated to them.

The order in which participants took part in the Feedback and No-Feedback conditions were randomised to help reduce the order effects.

*4)* Outcome Measures

The outcome measures consisted mainly of measures related to the chewing behaviour (i.e. chewing rate) and the self-awareness of participants regarding their eating activity.

*i) Chewing Behaviour*

During each meal phase, information was recorded in real time including the onset and termination of each individual eating event. A number of variables were extracted that were hypothesised as potentially affected by feedback, including: chewing rate across the entire eating phase, repeated measures of chewing rate across an eating sequence, the duration of detected events, and the period between detected events.

The live chewing rate was calculated and recorded using Eq. (2). However, this rate was sensitive to pauses between mouthfuls of food and as such was not used as an accurate indicator of chewing rate whilst eating across the entire meal phase. Instead, during data analysis, substantial gaps between chewing events were considered an indication of a pause following completion of a chewing sequence, or mouthful of food. During such a pause, a participant would swallow food and take in another portion for processing. Based on this, an adjusted chewing rate could be calculated to compensate for such pauses, by attenuating periods between chewing events which exceeded a given threshold. In this way, corrected values were found for chew event onset, $corr_{on}$, and termination, $corr_{off}$ (see Eqs. (3-5)). Simultaneously, this process could be used to identify the onset of chewing sequences, $seq_{on}$, and termination times of chewing sequences, $seq_{off}$.

Once these corrected times were found, the chewing rate over the whole session, $CR_{overall}$, could be defined as the number of detected chew events, $L$, divided by the time, in seconds, between the onset of the first chew event and termination of the last. It was calculated as follows:

$$\text{CR\_overall} = \frac{1}{L}(\text{corr\_off}_L - \text{corr\_on}_0) \quad (3)$$

Additional measures of eating were derived from the detected eating events. These measures included: average duration of chewing events, average period between chewing events, average duration of chewing sequences, average period between chewing sequences, and average number of chews per chewing sequence. Average duration of chewing events, $chew_{dur}$, was determined by the following equation:

$$\text{chew\_dur} = \frac{1}{L}\sum_{i=0}^{L}(\text{chew\_off}_i - \text{chew\_on}_i) \quad (4)$$

The average period between chewing events, chew$_{gap}$, was determined by the following equation:

$$\text{chew\_gap} = \frac{1}{L}\sum_{i=0}^{L}(\text{corr}_{\text{on}_i} - \text{corr}_{\text{off}_{i-1}}) \quad (5)$$

Following identification of chewing sequences based on a threshold for identifying significant gaps between chewing events, as discussed previously, the duration of and period between chewing sequences could be calculated. For instance, given the identification of chewing sequence onset (seq$_{on}$) and chewing sequence termination (seq$_{off}$), the average duration of eating sequences, seq$_{dur}$, and average period between chewing sequences, seq$_{gap}$, per meal could be calculated.

*ii) Self-awareness in the eating activity*

A short survey was administered after each phase of the study to gauge the participants' self-awareness of their eating activity and awareness of the food being consumed during that phase. The concept of 'mindfulness' of one's eating activity has been suggested as an important factor in supporting eating behaviour change and countering automated eating behaviour as a result of environmental factors [3, 33]. In order to measure such effects, previous studies [2] had employed surveys such as the "Kentucky Inventory of Mindfulness" to capture participants' degree of mindfulness in day-to-day life [34], and the "Three Factor Eating Questionnaire" to identify participants' dietary restraint, disinhibition and hunger in a general context [35]. Whilst these give a general context of participant mindfulness and eating behaviour, they do not provide details regarding participant mindfulness or eating behaviour in regards to a particular task, or during said task. Based on these questionnaires, a custom questionnaire was developed which consisted of 23 statements regarding participant self awareness of eating, rated on a 5 point Likert-scale from 'strongly disagree' to 'strongly agree'. The statements were selected in an attempt to provide insights into participant awareness of their environment, eating behaviour, eating speed and their overall self-awareness. Participants were asked to consider a normal eating scenario and compare their experience with recently completed eating phase, then score each statement. The survey responses were numerically categorised, between 1= "strongly disagree" and 5= "strongly agree". For each factor, an awareness score was defined as an average of all responses for statements related to that factor.

*F. Experimental Results*

Repeated measures of Analysis of Variance (ANOVA) were carried out to investigate differences between the 6 measures used to evaluate the chewing behaviour of participants between the control, No-Feedback and Feedback conditions (Total chewing rate, Chewing sequence duration, Chewing event duration, Time between the chewing event, Time between the chewing sequence and the Number of chewing events per sequence). Shapiro-Wilk statistical tests were used to check for normality. In a handful of the variables, the tests indicated that data deviated from normal distributions. Hence we ran non-parametric tests for such cases, which gave functionally the same results. See S2, S3 and S4 in supplementary materials for the full analysis (including normality test). The assumption of sphericity was tested using Mauchlys Test of sphericity, where this test showed that sphericity was violated Greenhouse-Geisser correction for violations of sphericity was used, otherwise sphericity was assumed. Table 5 shows a summary of the results.

TABLE 5
A SUMMARY OF THE DIFFERENCES BETWEEN THE MEASURES USED TO EXAMINE CHEWING BEHAVIOR

| Measurement (Mean Value) | Condition | | |
|---|---|---|---|
| | Control | No Feedback | Haptic Feedback |
| Total Chewing rate* | 1.6 (SD=0.32) | 1.18 (SD=0.34) | 0.92 (SD=0.35) |
| Chewing sequence duration* | 4.84 (SD=0.92) | 5.48 (SD=1.31) | 7.64 (SD=2.16) |
| Chewing event duration* | 0.42 (SD=0.07) | 0.48 (SD=0.11) | 0.53 (SD=0.15) |
| Time between Chewing event* | 0.34 (SD=0.16) | 0.59 (SD=0.23) | 0.86 (SD=0.45) |
| Time between Chewing sequence* | 1.56 (SD=0.65) | 1.86 (SD=0.60) | 2.72 (SD=1.31) |
| Number of chewing events per sequence* | 6.50 (SD=0.77) | 5.39 (SD=0.85) | 6.03 (SD=1.12) |

* indicates a statistically significant difference (p<0.001).

Overall, the repeated ANOVAs showed that there was a statistically significant difference (p<0.01) between the 3 conditions with regards to Total chewing rate (F (2,38) = 58.243, p <0.001), Chewing sequence duration ( F(2,38) = 31.696, p <0.001), Chewing event duration ( F(2,38) = 5.843, p<0.01), Time between chewing sequence (F(1.3, 24.7) = 16.65, p< 0.001 ), Time between chewing event ( F(2,38) = 66.01, p<0.001) and chew events per chewing sequence (F(1.52,28.99) = 9.78, p<0.01). Post-hoc tests showed that there was a statistically significant difference between the (haptic) Feedback condition and the No-(haptic)Feedback and control conditions, the results of which are summarised in Table 6.

TABLE 6
POST-HOC COMPARISON SHOWING THE DIFFERENCES IN THE MEASURES RELATED TO CHEWING BEHAVIOR BETWEEN THE DIFFERENT CONDITIONS

| Measurement | | Mean Difference (I - J) | Std. Error | p-value |
|---|---|---|---|---|
| Total Chewing rate | Control - No Feedback* | 0.425 | 0.051 | **0.000** |
| | Control - Haptic Feedback* | 0.676 | 0.063 | **0.000** |
| | No Feedback - Haptic Feedback* | 0.251 | 0.054 | **0.001** |
| Chewing sequence duration | Control - No Feedback* | -0.641 | 0.216 | 0.024 |
| | Control - Haptic Feedback* | -2.807 | 0.465 | **0.000** |
| | No Feedback - Haptic Feedback* | -2.166 | 0.401 | **0.000** |
| Chewing event duration | Control - No Feedback | -0.065 | 0.030 | 0.160 |
| | Control - Haptic Feedback* | -0.111 | 0.034 | **0.012** |
| | No Feedback - Haptic Feedback | -0.048 | 0.030 | 0.390 |
| Time between Chewing event | Control - No Feedback | -0.302 | 0.123 | 0.073 |
| | Control - Haptic Feedback* | -1.161 | 0.306 | **0.004** |
| | No Feedback - Haptic Feedback* | -0.859 | 0.231 | **0.004** |
| Time between Chewing sequence | Control - No Feedback* | -0.249 | 0.038 | **0.000** |
| | Control - Haptic Feedback* | -0.521 | 0.089 | **0.000** |
| | No Feedback - Haptic Feedback* | -0.272 | 0.080 | **0.009** |
| Number of chewing events per sequence | Control - No Feedback* | 1.107 | 0.207 | **0.000** |
| | Control - Haptic Feedback | 0.473 | 0.245 | 0.207 |
| | No Feedback - Haptic Feedback | 0.634 | 0.323 | 0.193 |

Overall, the results showed that on average, the lowest observed chewing rate was found in the Feedback condition,

(Mean= 0.92, SD=0.35) which was significantly lower than the No-Feedback condition (M=1.18, SD=0.34) and the control condition (Mean=1.6, SD=0.32). Participants in the Feedback condition showed on average, the longest chewing duration when consuming their food in the Feedback condition (Mean=7.64, SD=2.16), which was significantly longer than the No-Feedback condition (Mean=5.48, SD=1.31) and the control condition (Mean=4.84, SD=0.92). On average, participants spent significantly more time chewing in the Feedback condition (Mean=0.53, SD=0.15) than the control condition (Mean=0.42, SD=0.07). However, there was not a significant difference in the average chewing event duration between the Feedback and No-Feedback condition and the No-feedback and control condition.

Participants in the haptic Feedback condition spent on average significantly more time between each chewing event (Mean=0.86, SD =0.45) than No-Feedback (Mean=0.59, SD=0.23) and control condition (Mean=0.34, SD=0.16). Similarly, on average participants spent significantly more time between each chewing sequence when provided in the Feedback condition (Mean=2.72, SD=1.31) than the No-Feedback condition (Mean=1.86, SD=0.60) and the control condition (Mean= 1.56, SD=0.65). Finally, whilst participants in the control condition showed on average, a significantly higher number of chewing events per each chewing sequence (Mean=6.5, SD=0.77) than the No-Feedback condition (Mean= 5.39, SD= 0.85), there was not a significant difference in the number of chewing events per sequence in the Feedback and Non-Feedback and the Feedback and control conditions

TABLE 7
POST-HOC COMPARISON SHOWING THE DIFFERENCES IN THE MEASURES RELATED TO SELF AWARENESS MEASURES.

| Measurement | | Mean Difference (I - J) | P-value |
|---|---|---|---|
| Environment | Control - No Feedback | 0.400 | 0.159 |
| | Control - Haptic Feedback* | 0.438 | **0.034** |
| | No Feedback - Haptic Feedback | 0.038 | 1.0 |
| Eating | Control - No Feedback | -0.240 | 0.161 |
| | Control - Haptic Feedback* | -0.317 | **0.008** |
| | No Feedback - Haptic Feedback | -0.077 | 1.0 |
| Speed | Control - No Feedback* | -0.925 | **0.000** |
| | Control - Haptic Feedback* | -1.037 | **0.000** |
| | No Feedback - Haptic Feedback | -0.113 | 1.0 |

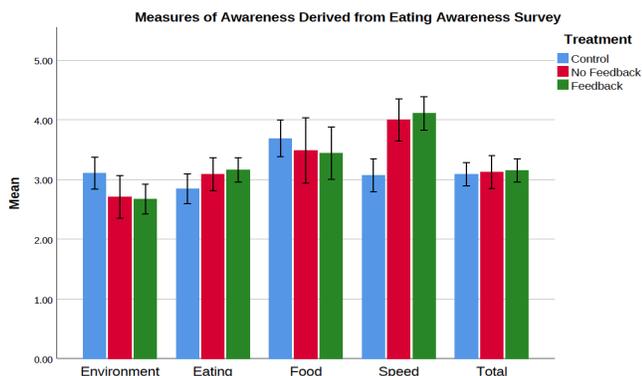

Figure 4: The means score of each of the measures for self-awareness factors examined in this study

For measures of awareness, the same procedure was used for analysing awareness factors as for chewing measures. A repeated measure ANOVA was carried out with Bonferroni adjusted post-hoc tests to determine differences between conditions (see S5-S7 in the Supplementary Material for full analysis). In this case, all factors were found to have a normal distribution. For the environment factor, post-hoc pairwise comparison indicated that there was a significant difference between the control condition and haptic Feedback condition in the participants' awareness in their environment ($p < 0.05$). For the factor of awareness with regards to their eating activity, the score of participants during the Feedback condition was found to be significantly higher than the control condition ($p < 0.01$). Finally, participants reported significantly higher awareness of their eating speed during both the Feedback condition and the No-Feedback condition than during the control period ($p < 0.001$). Figure 4 shows a summary of the average score and standard deviation for each of the self-awareness factors measured in this study. The results of the post-hoc analysis for factors which showed significant differences in the ANOVA tests are summarised in Table 7.

## V. DISCUSSION

The paper reported two studies; study 1 focused on the development of an algorithm to detect eating behaviour, whilst study 2 presented an experimental study looking into the use of haptic feedback to facilitate mindful eating. Using EMG signals of the masseter and submental muscles, our classifier algorithms based on a linear SVM, was capable of swallow detection with an accuracy of 87% and chew detection with an accuracy of 94%. In addition, the algorithm was shown to be robust and able to generalise well in a leave-one-participant-out evaluation scheme. This was achieved through use of data from 16 participants over a wide range of BMI values, and including natural behaviour aspects such as head motion, reading aloud, etc. In the second study, we showed through an experiment with 20 participants, that haptic feedback triggered by automatic eating behaviour detection, had a significant affect in supporting voluntary eating rate reduction; resulting in a significant difference in eating rate between treatment groups, with an average rate during feedback based moderation 46.9% slower than the no-feedback moderation. These studies demonstrated the use of eating driven real-time feedback for the purpose of behaviour change intervention through providing ongoing reminders of chewing moderation goals.

### A. Discussion on the classification

The first goal of this study was to develop classifier algorithms for automated chewing and swallowing detection based on EMG signals. Overall, we found that the models were robust, generalising well across different BMI and age range. Compared to previous studies, the results of the study reported here were accurate in the presence of unrelated activities (e.g. reading, head motions, etc.). For instance, 'smart-glasses' based studies showed comparable performance for chewing detection using threshold based algorithms [36, 37]. Huang et al. [36] reported an accuracy of 96%, however they indicated a high degree of false positives associated with unexpected activity. Similarly, Zhang and Amft [37] reported chewing detection accuracy of approximately 94% for their algorithm in lab conditions, but only 80% accuracy in a more realistic settings. Our swallowing detection classifier resulted in an accuracy of 87% (F1-score=0.87), which was lower than the accuracy of 93% reported by Nahrstaedt et al. [38] using a combined bioimpedance and EMG based algorithm. However, the higher performance in Nahrstaedt et al. [38] might be attributable to a number of factors, such as a limited subject pool, consisting of 9 subjects, two of whom were female (mean age 28.5), and seven male (mean age 27.4), with unspecified BMI differences. The study also involved experimentally controlled bolus size swallowed, and different sensor placement, across the sternohyoid muscle rather than submental muscles. Furthermore, the inclusion of both bioimpedance and EMG may add additional processing costs to the detection of swallowing activity, while the approach proposed in this study relies solely upon analysis of a single EMG channel.

### B. Discussions on the mindful eating experiment

Eating speed and chewing thoroughness have been suggested as factors impacting various aspects of physical health such as increasing the possibility of a high BMI or increasing the risk of developing eating disorders [39]. In this study, we carried out an experiment to investigate the potential of an automated chew monitoring system with haptic feedback to help participants moderate their chewing rate. From the experiment, we found that participants exhibited a lower rate of chewing during self-moderation of eating than during normal eating condition and were found to further reduce chewing rate through the use of haptic feedback. Overall, we found significant increase in the period between chews in the feedback condition compared with the control, which was again larger in the haptic feedback condition. Interestingly, no statistically significant difference (p=0.39) was found in the duration of chewing events between the no feedback and haptic feedback conditions. This suggests that although participants spent longer chewing each mouthful during moderation, particularly when supported by haptic feedback, the duration of individual chews remained relatively constant. The average number of chewing events occurring during each chewing sequence could be considered as an indication of chewing thoroughness. Like the chewing event duration, for this measure there was no significant differences. The average number of chews per chewing sequence remained relatively constant. Furthermore, the lack of change in number of chews or duration of chewing events implies that the increase in average duration of chewing sequences, and chewing rate in general, may primarily be a function of the time between individual chews rather than other factors.

In the experiment, participants' self-awareness was estimated from Likert scale type responses to a number of statements to estimate overall levels of mindfulness related to eating. Mindful eating has been suggested as a component of eating behaviour change [2, 34, 33] and it was hypothesised here that self-moderation and feedback would have an impact upon participants self-awareness regarding eating. Our results only partially supported this hypothesis. No difference was identified between the conditions for participant awareness scores focusing upon food (p=0.71), or for total awareness (p=0.78). However, statistically significant differences were found for participant's self-awareness in relation to their environment, eating behaviour, and in regards to their focus upon eating speed (p<0.01). Although participants appeared to be more aware of their eating environment during the control condition. Whilst counterbalancing was applied between the No-Feedback and Feedback conditions to moderate any temporal effects, the control condition was always carried out prior to these. This was done to enable calibration of the system and for baseline measurement. As such, there is a potential that differences between control and treatment periods was the result of participants becoming familiar with the setting, and less self-aware regarding their environment. This may also explain the effect upon eating awareness and participant's awareness on their speed of eating. The scores for eating speed awareness were higher during the non-feedback and haptic feedback than the control, but did not differ significantly between one another.

### C. Implications

The detection of various eating related features may be useful for providing valuable health-related feedback. In addition to visual evaluation of health (for instance through EMG for swallowing function monitoring), feedback regarding physiological processes and physical activity has been used for treatment of certain health conditions. For example, biofeedback aims to help an individual gain voluntary control of physiological processes to help treat conditions, as part of rehabilitation following a stroke [40], or for helping practice swallowing rehabilitation exercises in the treatment of swallowing disorders [41]. The technological approach we developed has the potential for other applications, for example, providing daily feedback regarding dietary intake goals based on automated detection of intake technique which has been used in conjunction with mobile based self-report of diet for weight change goals [42].

Previous studies had highlighted that the mobility and popularity of mobile devices, along with potential for personalised feedback and goal management, may facilitate tracking dietary intake, exercise or weight management, and eating related interventions [42, 24]. In particular, mobile phones could be particularly useful in automated systems for dietary tracking, eating monitoring, or for goal based intervention or therapy, as a way to provide feedback, permit goal setting, and review of progress. Thus, the technology developed in this study could be particularly useful in weight change interventions: for providing feedback, encouraging the adoption of eating patterns and styles which have been associated with increased satiation and reduced intake [20], or for detecting adherence to a diet plan, using a model trained to detect specific foods. Such system might help support clinical diet change for treatment of obesity, or monitoring adherence to set diets prior to some surgeries or other treatment, sharing data regarding intake directly with medical staff. In regards to weight management, there are also implications of the system developed in this paper for the screening and monitoring of eating disorders during treatment. Traditionally, screening of eating disorders is carried out subjectively through clinical interviews and questionnaires. The classification models developed here, in conjunction with intake volume estimation and data sharing can be of considerable benefit to eating disorder treatment. Eating activity might be evaluated to identify patterns which are characteristic of eating disorders, such as periods of fasting, binging [43], or event related to eating speed [10]. Potentially, compensatory activities might also be detected, such as purging, based on facial muscle activity, or excessive exercise through the use of additional sensors (such as exercise tracking bands).

## VI. CONCLUSION

We presented a novel system for automatically detecting eating behaviour in real time using EMG sensing. We demonstrated the use of wearable haptic feedback device to help facilitate mindful eating. Overall, the work carried out in this study has major implications for several areas of research, particularly for studying eating habits and improving our understanding of eating behaviour and the various influences upon eating choices such as food selection, intake volume, and intake speed. Automated eating detection systems may instead permit accurate collection of information with comparatively minimal processing. The methodology developed for the detection of eating speed could be extended to other forms of feedback regarding the users' eating rate (audio, visual, and haptic). The impact of different distraction types (television, music, or other stimuli), social meals, and portion sizes upon eating speed, or the effect of feedback or different stress conditions, might all be investigated using the system developed in this paper, with appropriate adaptation. Finally, eating speed might be investigated across demographic groups, to determine any particular associations between individuals with differing BMI, obesity, diabetes, or different eating disorders.

BIO

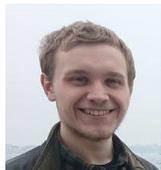 **Benjamin Peter Parry Nicholls** is a Doctoral candidate at the University of Kent finalising his thesis. His doctoral research focuses on using physiological sensing to remove the burden and error inherent in eating self-report. He is interested in mobile and discrete on-body sensing technologies, machine learning for monitoring of eating function and dietary content, and applications of eating related feedback. He also holds a masters degree in Computer Science and is currently working as a Data Scientist in the social and financial data sector. Ben was born in Canterbury, Kent, and outside of research enjoys reading, the outdoors, and tabletop games.

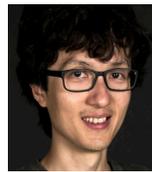 **Chee Siang Ang** is a Senior Lecturer in Multimedia and Digital Systems in the School of Engineering and Digital Arts, University of Kent. Before joining Kent, he was a research fellow at the Centre for Human Computer Interaction Design, City University London, where he completed his PhD in the area of social gaming. His main research area is in digital health, where he investigates, designs and develops new technologies which can provide treatment and (self-) management of health conditions, through effective prevention, early intervention, personalised treatment and continuous monitoring of the conditions. He is particularly interested in immersive media technologies (virtual or augmented reality), computer games, sensing technologies.

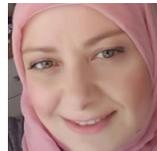 **Eiman Kanjo**, is an Associate Professor in Mobile Sensing & Pervasive Computing at Nottingham Trent University. She is a technologist, developer and an active researcher in the area of mobile sensing, smart cities, spatial analysis, and data analytics, who worked previously at the University of Cambridge, Mixed Reality Laboratory and the University of Nottingham as well as the International Centre for Computer Games and Virtual Entertainment, Dundee. She has authored some of the earliest papers in the research area of Mobile Sensing and currently carries out work in the area of Digital Phenotyping Smart cities, technologies for Mental Health and the Internet of Things for Behaviour Change.

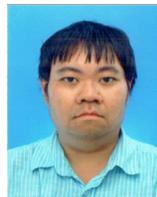 **Panote Siriaraya** is an Assistant Professor at the Faculty of Information and Human Science, Kyoto Institute of Technology in Japan. He received his PhD degree in Electronics from the University of Kent in Canterbury, UK in 2013. Afterwards, he worked as a post-doc researcher at the Faculty of Industrial Design Engineering at the Delft University of Technology from 2014 to 2017. His main research interest is in the field of Human-Computer Interaction, which includes topics such as Virtual Environments, Gamification and Designing Technologies for the Aging population.

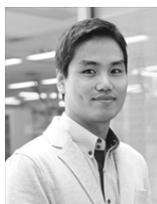 **Woon-Hong Yeo** is an Assistant Professor in the George W. Woodruff School of Mechanical Engineering and the Wallace H. Coulter Department of Biomedical Engineering at Georgia Institute of Technology. He received his B.S. in mechanical engineering from Inha University in 2003 and his PhD in mechanical engineering at the University of Washington, Seattle in 2011. From 2011-2013, he worked as a postdoctoral research fellow at the Beckman Institute and Frederick Seitz Materials Research Center at the University of Illinois at Urbana-Champaign. His research areas include soft electronics, human-machine interfaces, nano-biosensors, and soft robotics (research website: http://yeolab.gatech.edu).

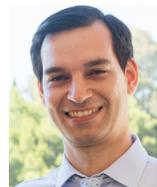 **Athanasios Tsanas ('Thanasis')** (MIEEE) completed a BSc in Biomedical Engineering at the Technological Educational Institute of Athens (2005), a BEng in Electrical Engineering and Electronics at the University of Liverpool (2007), an MSc in Signal Processing and Communications at the Newcastle University (2008), and a DPhil (PhD) in Applied Mathematics, Mathematical Institute, University of Oxford (2012). He was previously a post-doctoral research fellow at the University of Oxford (2012-2016) and a Stipendiary Lecturer in Engineering Science (2014-2016). He is currently an Assistant/Associate Prof. in Data Science (tenure-track) at the

University of Edinburgh, Edinburgh UK, and a Lecturer in Statistical Research Methods at the University of Oxford. Dr. Tsanas was the recipient of the Andrew Goudie award (top PhD student across all disciplines, St. Cross College, University of Oxford, 2011), the EPSRC Doctoral Prize award (2012), a winning team member in the Computing in Cardiology annual competition for 'predicting mortality in the ICU' (2012), the young scientist award (MAVEBA, 2013), the EPSRC Statistics and Machine Learning award (2015), and won a 'Best reviewer award' from the IEEE Journal of Biomedical Health Informatics (2015). He is sitting at the Editorial Board of JMIR mHealth and uHealth, and the Editorial Board of JMIR Mental Health.

# Supplementary Material

## S1: 18 Features extracted from EMG Signals

| Feature name | Method |
|---|---|
| Mean Absolute Value (MAV) [1], [2], [3], [4], [5], [6], [7], [8] | Average of the absolute EMG signal across a signal segment. Defined as: $$\mu = \frac{1}{N}\sum_{i=1}^{N}|x_i|$$ Where $x_i$ is the EMG signal sampled at time $i$ and $N$ is the number of samples |
| Integrated EMG (IEMG) [1], [2], [9], [10], [11] | Related to EMG signal firing point. Defined as the summation of the absolute EMG signal across an EMG segment $$IEMG = \sum_{i=1}^{N}|x_i|$$ Where $x_i$ is the EMG signal sampled at time $i$ and $N$ is the number of samples |
| Variance (VAR) [1], [2] | Variance of EMG signal across a segment $$VAR = \frac{1}{N-1}\sum_{i=1}^{N}(x_i^2 - \bar{x})$$ Where $\bar{x}$ is the mean of the segment, $x_i$ is the EMG signal sampled at time $i$ and $N$ is the number of samples |
| Root Mean Square (RMS) [1], [2], [12], [3] | Square root of the average square of EMG amplitude across a segment $$RMS = \sqrt{\frac{1}{N}\sum_{i=1}^{N}x_i^2}$$ Where $x_i$ is the EMG signal sampled at time $i$ and $N$ is the number of samples |
| Standard Deviation (SD) [1], [2], [3] | Standard deviation ($\sigma$) of the EMG signal across a given segment of EMG signal: $$\sigma = \sqrt{\frac{1}{N-1}\sum_{i=1}^{N}(x_i - \bar{x})^2}$$ Where $\bar{x}$ is the mean of the segment, $x_i$ is the EMG signal sampled at time *i* and *N* is the number of samples in the segment |
| Waveform Length (WL) [1], [2] | Cumulative length of EMG waveform over a signal segment $$\text{WL} = \sum_{i=1}^{N-1}|x_{i+1} - x_i|$$ Where $x_i$ is the EMG signal sampled at time *i* and *N* is the number of samples |
| Peak Amplitude [13], [4], [7], [5], [6] [8] | The peak amplitude across a given segment of the EMG signal |
| Myopulse Percentage Rate [1], [2] | Related to the firing of Motor Unit Action Potentials. Average Number of times that the absolute of the EMG signal exceeds *thr* $$MYOP = \frac{1}{N}\sum_{i=1}^{N}[f(|x_i|)], f(x) = \begin{cases} 1, & x \geq thr \\ 0, & x < thr \end{cases}$$ Where $x_i$ is the EMG signal sampled at time *i* and *N* is the number of samples |
| Willison Amplitude (WAMP) [1], [2] | Sum of times the absolute EMG exceeds a given threshold *thr*: $$WAMP = \frac{1}{N-1}\sum_{i=1}^{N}[f(|x_i - x_{i+1}|)], f(x) = \begin{cases} 1, & x \geq thr \\ 0, & x < thr \end{cases}$$ Where $x_i$ is the EMG signal sampled at time *i* and *N* is the number of samples |

| Zero crossing (ZC) [1], [2] | Number of times EMG amplitude crosses zero amplitude: $$ZC = \frac{1}{N-1}\sum_{i=1}^{N}\left[sgn\left((x_i \times x_{i+1}) \cap |x_i - x_{i+1}|\right) \geq thr\right]$$ $$sgn = \begin{cases} 1, & x \geq thr \\ 0, & x < thr \end{cases}$$ Where $x_i$ is the EMG signal sampled at time $I$, $N$ is the number of samples, and $thr$ is a predefined crossing threshold |
|---|---|
| Slope Sign Change (SSC) [1], [2] | Count of the number of times the EMG signal slope changes across a signal segment $$\frac{1}{N-1}\sum_{i=1}^{N}\left[f\left((x_i - x_{i-1}) \times (x_i - x_{i+1})\right)\right]$$ $$f(x) = \begin{cases} 1, & x \geq thr \\ 0, & x < thr \end{cases}$$ Where $x_i$ is the EMG signal sampled at time $i$ and $N$ is the number of samples |
| Mean Frequency (MNF) [1], [2], [12] | Average frequency calculated by: $$\text{MNF} = \frac{\sum_{j=1}^{M} f_j P_j}{\sum_{j=1}^{M} P_j}$$ Where $f_j$ is the frequency of the power spectrum at frequency $j$, $P_j$ is the EMG power spectrum at frequency bin $j$ and $M$ is the length of the frequency bin. |
| Mean Power Spectrum (MNP) [1], [2], [12] | Average power spectrum of the EMG signal sample: $$MNP = \frac{1}{M}\sum_{j=1}^{M} P_j$$ Where $P_j$ is the EMG power spectrum at frequency bin $j$ and $M$ is the length of the whole frequency bin. |
| Median Frequency (MDF) [1], [2], [12] | Frequency at which the spectrum is divided into two regions of equal amplitude $$\sum_{j=1}^{MDF} P_j = \sum_{j=MDF}^{M} P_j = \frac{1}{2}\sum_{j=1}^{M} P_j$$ Where $P_j$ is the EMG power spectrum at frequency bin $j$ and $M$ is the length of the whole frequency bin. |
| Median Power Frequency (MPF) [1], [2], [12] | Band power of the median frequency calculated using Fast Fourier Transform |
| $T_p$ values [13], [14], [15] | Defined as the normalized time point across a chewing cycle at which point $P$ percent of the total cumulative EMG has occurred. Calculated using the following steps. 1. Calculate cumulative sum across sample window 2. Normalised duration of sample 3. $T_p$ is the normalized time at which $P$ percent of the cumulative sum of the signal has occurred |
| Cycle Duration [13], [16] [5], [6], [4], [7], [8] | Duration of a chew or swallow EMG activity cycle from onset to termination |
| Cycles per sequence [13], [15], [17] | Count of the number of chewing cycles within a given chewing sequence |

## S2: Shapiro-Wilk normality test for chewing measures

| Eating Measures: Normality Test | | | | |
|---|---|---|---|---|
| Treatment | Measure | Shapiro-Wilk | | |
| | | Statistic | df | Sig. |
| Control | Chewing Rate | 0.986 | 20 | 0.987 |
| | Chewing Sequence Duration | 0.823 | 20 | 0.002 |
| | Chewing Event Duration | 0.888 | 20 | 0.025 |
| | Time Between Chew Events | 0.868 | 20 | 0.011 |
| | Time Between Chewing Sequence | 0.954 | 20 | 0.424 |
| | Chew Cycle per Chewing Sequence | 0.963 | 20 | 0.600 |
| No feedback | Chewing Rate | 0.927 | 20 | 0.136 |
| | Chewing Sequence Duration | 0.904 | 20 | 0.049 |
| | Chewing Event Duration | 0.813 | 20 | 0.001 |
| | Time Between Chew Events | 0.962 | 20 | 0.594 |
| | Time Between Chewing Sequence | 0.978 | 20 | 0.900 |
| | Chew Cycle per Chewing Sequence | 0.954 | 20 | 0.428 |
| Feedback | Chewing Rate | 0.891 | 20 | 0.028 |
| | Chewing Sequence Duration | 0.909 | 20 | 0.061 |
| | Chewing Event Duration | 0.851 | 20 | 0.006 |
| | Time Between Chew Events | 0.896 | 20 | 0.035 |
| | Time Between Chewing Sequence | 0.937 | 20 | 0.211 |
| | Chew Cycle per Chewing Sequence | 0.871 | 20 | 0.012 |

# S3: Repeated measure Analysis of Variance to determine significant differences between treatments/conditions, for chewing measures

| | Chewing Measure: Repeated Measure ANOVA Test | | | | | |
|---|---|---|---|---|---|---|
| | | df | Mean Square | F | Sig. | Partial Eta Squared |
| Treatment | Total Chew Rate | 2 | 0.330 | 58.243 | 0.000 | 0.754 |
| | Chew Sequence Duration | 2 | 0.188 | 31.696 | 0.000 | 0.625 |
| | Chew Event Duration | 2 | 0.045 | 5.843 | 0.006 | 0.235 |
| | Period Between Chew Sequences* | 1.304 | 0.400 | 16.645 | 0.000 | 0.467 |
| | Period Between Chew Cycles | 2 | 0.786 | 66.007 | 0.000 | 0.776 |
| | Chew Events per Chew Sequence* | 1.525 | 0.046 | 9.775 | 0.001 | 0.340 |
| Error (Treatment) | Total Chew Rate | 38 | 0.006 | | | |
| | Chew Sequence Duration | 38 | 0.006 | | | |
| | Chew Event Duration | 38 | 0.008 | | | |
| | Period Between Chew Sequences* | 24.777 | 0.024 | | | |
| | Period Between Chew Cycles | 38 | 0.012 | | | |
| | Chew Events per Chew Sequence* | 28.981 | 0.005 | | | |

*. Sphericity Violated Greenhouse-Geisser determines significance

# S4. Post-hoc comparison of chewing measures for each treatment /condition

| Chewing Measure: Post-Hoc Pairwise Comparisons | | | | | | | |
|---|---|---|---|---|---|---|---|
| Measure | | | Mean Difference (I-J) | Std. Error | Sig.[b] | 95% Confidence Interval for Difference[b] | |
| | | | | | | Lower Bound | Upper Bound |
| Total Chewing Rate | Control | No Feedback | 0.425* | 0.051 | 0.000 | 0.292 | 0.558 |
| | | Feedback | 0.676* | 0.063 | 0.000 | 0.510 | 0.842 |
| | 'No Feedback | Feedback | 0.251* | 0.054 | 0.001 | 0.109 | 0.393 |
| Duration of Chew Sequence | Control | No Feedback | -0.641* | 0.216 | 0.024 | -1.207 | -0.074 |
| | | Feedback | -2.807* | 0.465 | 0.000 | -4.027 | -1.586 |
| | 'No Feedback | Feedback | -2.166* | 0.401 | 0.000 | -3.219 | -1.114 |
| Duration of Chew Event | Control | No Feedback | -0.062 | 0.030 | 0.160 | -0.142 | 0.017 |
| | | Feedback | -0.111* | 0.034 | 0.012 | -0.199 | -0.022 |
| | 'No Feedback | Feedback | -0.048 | 0.030 | 0.390 | -0.128 | 0.032 |
| Period Between Chew Sequence | Control | No Feedback | -0.302 | 0.123 | 0.073 | -0.626 | 0.022 |
| | | Feedback | -1.161* | 0.306 | 0.004 | -1.965 | -0.358 |
| | 'No Feedback | Feedback | -0.859* | 0.231 | 0.004 | -1.465 | -0.253 |
| Period Between Chew Event | Control | No Feedback | -0.249* | 0.038 | 0.000 | -0.349 | -0.149 |
| | | Feedback | -0.521* | 0.089 | 0.000 | -0.754 | -0.288 |
| | 'No Feedback | Feedback | -0.272* | 0.080 | 0.009 | -0.482 | -0.062 |
| Chew Events per Sequence | Control | No Feedback | 1.107* | 0.207 | 0.000 | 0.563 | 1.651 |
| | | Feedback | 0.473 | 0.245 | 0.207 | -0.171 | 1.117 |
| | 'No Feedback | Feedback | -0.634 | 0.323 | 0.193 | -1.482 | 0.213 |

Based on estimated marginal means

*. The mean difference is significant at the .05 level.

[b]. Adjustment for multiple comparisons: Bonferroni.

# S5: Shapiro-Wilk normality test for awareness measures

| | Awareness: Normality Test | | | |
|---|---|---|---|---|
| Treatment | Awareness Factor | Statistic | df | Sig. |
| Control | Environment | 0.953 | 20 | 0.407 |
| | Eating | 0.971 | 20 | 0.769 |
| | Food | 0.970 | 20 | 0.754 |
| | Speed | 0.969 | 20 | 0.735 |
| | Total | 0.965 | 20 | 0.649 |
| No Feedback | Environment | 0.902 | 20 | 0.050 |
| | Eating | 0.960 | 20 | 0.537 |
| | Food | 0.917 | 20 | 0.086 |
| | Speed | 0.918 | 20 | 0.091 |
| | Total | 0.923 | 20 | 0.112 |
| Feedback | Environment | 0.926 | 20 | 0.131 |
| | Eating | 0.932 | 20 | 0.168 |
| | Food | 0.835 | 20 | 0.053 |
| | Speed | 0.916 | 20 | 0.082 |
| | Total | 0.938 | 20 | 0.224 |

## S6: Repeated measure Analysis of Variance to determine significant differences between treatments/conditions, for awareness measures

| | | Overall Awareness: Repeated Measure ANOVA | | | | | |
|---|---|---|---|---|---|---|---|
| | Factor | Type III Sum of Squares | df | Mean Square | F | Sig. | Partial Eta Squared |
| Treatment | Environment | 2.352 | 2 | 1.176 | 3.964 | 0.027 | 0.173 |
| | Eating | 1.092 | 2 | 0.546 | 4.291 | 0.021 | 0.184 |
| | Food | 0.459 | 2 | 0.230 | 0.353 | 0.705 | 0.018 |
| | Speed | 12.965 | 2 | 6.482 | 30.815 | 0.000 | 0.619 |
| | Total | 0.065 | 2 | 0.032 | 0.246 | 0.783 | 0.013 |
| Error (Treatment) | Environment | 11.273 | 38 | 0.297 | | | |
| | Eating | 4.834 | 38 | 0.127 | | | |
| | Food | 24.744 | 38 | 0.651 | | | |
| | Speed | 7.994 | 38 | 0.210 | | | |
| | Total | 5.002 | 38 | 0.132 | | | |

# S7. Post-hoc comparison of awareness measures for each treatment /condition

| Overall Awareness: Post-Hoc Pairwise Comparisons | | | | | | | |
|---|---|---|---|---|---|---|---|
| Factor | | | Mean Difference (I-J) | Std. Error | Sig.[b] | 95% Confidence Interval for Difference[b] | |
| | | | | | | Lower Bound | Upper Bound |
| Environment | Control | No_Feedback | 0.400 | 0.194 | 0.159 | -0.109 | 0.909 |
| | | Feedback | 0.438* | 0.156 | 0.034 | 0.028 | 0.847 |
| | No-Feedback | Feedback | 0.038 | 0.165 | 1.000 | -0.394 | 0.469 |
| Eating | Control | No_Feedback | -0.240 | 0.117 | 0.161 | -0.546 | 0.066 |
| | | Feedback | -0.317* | 0.092 | 0.008 | -0.558 | -0.075 |
| | No-Feedback | Feedback | -0.077 | 0.127 | 1.000 | -0.410 | 0.256 |
| Food | Control | No_Feedback | 0.200 | 0.274 | 1.000 | -0.519 | 0.919 |
| | | Feedback | 0.167 | 0.221 | 1.000 | -0.413 | 0.747 |
| | No-Feedback | Feedback | -0.03 | 0.268 | 1.000 | -0.736 | 0.669 |
| Speed | Control | No_Feedback | -0.925* | 0.143 | 0.000 | -1.300 | -0.550 |
| | | Feedback | -1.037* | 0.145 | 0.000 | -1.419 | -0.656 |
| | No-Feedback | Feedback | -0.113 | 0.147 | 1.000 | -0.498 | 0.273 |
| Total | Control | No_Feedback | -0.032 | 0.123 | 1.000 | -0.354 | 0.290 |
| | | Feedback | -0.080 | 0.089 | 1.000 | -0.313 | 0.153 |
| | No-Feedback | Feedback | -0.048 | 0.129 | 1.000 | -0.386 | 0.290 |

Based on estimated marginal means
*. The mean difference is significant at the .05 level.
[b]. Adjustment for multiple comparisons: Bonferroni.